# Visible Photocatalytic Degradation of Methylene Blue on Magnetic Semiconducting La modified M-type Strontium Hexaferrite

Debesh Devadutta Mishra, Yao Huang, Guolong Tan[*]

State Key Laboratory of Advanced Technology for Materials Synthesis and Processing, Wuhan University of Technology, Wuhan 430070, China

**Abstract**

Methylene blue (MB) is a representative of a class of dyestuffs resistant to biodegradation. This paper presents a novel photocatalytic degradation of methylene blue (MB) by La modified M-type strontium hexaferrite ($La_{0.2}Sr_{0.7}Fe_{12}O_{19}$) compound, which is a traditional permanent magnet and displays a large magnetic hysteresis (M-H) loop. The remnant magnetic moment and coercive field are determined to be 52emu/g and 5876Oe, respectively. UV-Visible optical spectroscopy reveals that $La_{0.2}Sr_{0.7}Fe_{12}O_{19}$ is simultaneously a semiconductor, whose direct and indirect band gap energies are determined to be 1.47 eV and 0.88 eV, respectively. The near infrared band gap makes it a good candidate to harvest sunlight for photocatalytic reaction or solar cell devices. This magnetic compound demonstrates excellent photocatalytic activity on degradation of MB under visible illumination. The colour of MB dispersion solution changes from deep blue to pale white and the absorbance decreases rapidly from 1.8 down to zero when the illumination duration extends to 6 hours. Five absorption bands did not make any blue shifts along with the reaction time, suggesting a one-stepwise degradation process of MB, which makes $La_{0.2}Sr_{0.7}Fe_{12}O_{19}$ a unique magnetic catalyst and differs from $TiO_2$ and other conventional catalysts.

## 1. Introduction

As far back as Neolithic man first utilised a bit of suspended lodestone to explore, humankind has utilised attractive materials of different sorts. In any case, it was not until the approach of electricity that the magnetic processes started to be caught attraction. It is presently realised that lodestone is an iron mineral, magnetite, which is one of an extensive variety of magnetic ceramics, iron (III) oxide, called the ferrites having a cubic crystal structure. The ferrites or hard ferrites are also termed as oxide magnets have replaced older metal alloy permanent magnets and also its cost-effectiveness, lower weight and higher coercivity presented these materials as applications in dc motors electrical relays and ore separators.[1].

---

[*]Corresponding author; email: gltan@whut.edu.cn

The other group of ferrites, which have hexagonal structure, are known as hexaferrite. Hexaferrites are massively used in the commercial and technological application, such that Ba-hexaferrite provides 50% of magnetic materials with production is over 300,000 tonnes per year. Since the discovery, it is noticed that there has been an exponential increase in the degree of interest, due to general magnetic properties, microwave properties, data storage and magnetic recording materials. [2] There are various processes which were used to prepare hexaferrite such as calcination and sintering in high temperatures [3]. But this process produces coarser particulate sizes and is not economically viable. Some other processes such as coprecipitation, [4] spray-drying and microemulsion [5], hydrothermal [6], sol–gel [7, 8] and conventional route [9, 10]. Due to uniaxial magnetoplumbite structure with strontium divalent metallic ion, good magnetic properties, excellent tribological characteristics, weak temperature dependent coercivity and better chemical stability make $SrFe_{12}O_{19}$ a special material amongst hexagonal ferrites [11,12]. Various studies have been carried out in order to enhance the magnetic properties using substitutions on $Fe^{3+}$ or $Sr^{2+}$ ions or both with La–Co, $La^{3+}$, La–Zn, $Cr^{3+}$, and La-Co [13-19]. Amongst all the substitution, La addition is the most experimented upon strontium hexaferrite, whose magnetic performance was greatly enhanced. [20] Until now, the magnetic semiconducting characteristics have been tested upon several M-type hexaferrites, including $PbFe_{12-x}O_{19-\delta}$ [21] and Gd-Co doped barium hexaferrite [22].

But to the best of our knowledge, very few optical characterization studies have been carried out on La substituted on strontium hexaferrite. The photocatalytic performance of La substituted strontium hexaferrite upon its semiconducting feature has never been reported till to date. Titanium dioxide is the most widely used catalyst for pollutant degradation of photocatalytic process. [23-26] The band gap energy of $TiO_2$ is pretty large and it requires ultraviolet (UV) light to generate electrons for photocatalytic reaction. UV light makes up only a small fraction (<4%) of the total solar spectrum reaching the surface of the earth. [23] Therefore, seeking oxide semiconductors with band gap energies within visible light region to harvest most fraction of the visible sunlight would be a milestone for organic pollutant degradation through photocatalytic degradation process. La modified $SrFe_{12}O_{19}$ matches such requirements. The direct band gap energy locates at near infrared region, which makes it be able to harvest major sunlight of the solar spectrum. This compound contains only low toxic and abundance elements, and thus is environment friendly. The present study deals with the synthesis of $La_{0.2}Sr_{0.8}Fe_{12}O_{19}$ by a polymer precursor method. The semiconductor characterization of $La_{0.2}Sr_{0.8}Fe_{12}O_{19}$ has been checked out by a UV-visible-near infrared optical spectrometer, its excellent magnetic properties and photocatalytic role upon degradation process of methylene blue (MB) over the visible-light illumination have been investigated in detail.

## 2. Experimental Procedure

The preparation of La-$SrFe_{12}O_{19}$ has been carried out using polymer precursor procedure, where strontium acetate ($Sr(CH3COO)2 \cdot 3H_2O$), Lanthanum acetate ($La(CH3COO)2 \cdot 3H_2O$) and ferric acetylacetonate ($C_{15}H_{21}FeO_6$) were used (Alfa Aesar). At first stoichiometric compositions of strontium acetate and lanthanum acetate were dissolved in 15 ml glycerine and a clear solution is prepared. In order to remove the trapped water, the solution was distilled using a rotary evaporator at $120^0 C$. The distilled solution was transferred to the glove box in a 50 ml flask. A stoichiometric amount of ferric acetylacetonate was dissolved in a solution mixture containing 70 ml acetone and 100 ml of anhydrous ethyl alcohol in a three-necked flask inside the glove box. The above mixture was stirred for 6 hours at 70 ℃, in order to fully dissolve ferric acetylacetonate. The atomic ratio of strontium-lanthanum mixture to iron was kept to 1:9.5~10, in order to compensate the (Sr+La) loss during heat treatment process. Then the Fe precursor was mixed with Sr+La precursor to form a clear mixture solution.

Afterwards, 45 mL ammonia solution was poured into the above mixture solution. The dispersion solution was maintained at 70 ℃ under stirring for 24 hours and then moved out of the glove box. Through centrifugation (12000 rpm) of the dispersion solution, water and organic molecules were removed and the remnant colloidal substance was calcined at 450 ℃ for 1 hour. The calcined powders were grinded in an agate mortar for 1 hour and again heated at 800 ℃ for an hour to ensure the full removal of organic components. Finally the powders were air-sintered at 1100°C for 1 hour to ensure the formation of pure $La_{0.2}Sr_{0.7}Fe_{12}O_{19}$ in magnetoplumbite structure. Phase identification was characterised by means of X-ray diffraction (XRD) using a D8 RIGAKU X-ray diffractometer with Cu Kα. The magnetic properties of as-prepared powders was measured Using Quantum Design physical property measurement system (PPMS). The semiconducting feature of $La_{0.2}Sr_{0.7}Fe_{12}O_{19}$ compound has been evaluated by a Lambda 750 UV-Vis-near infrared optical spectroscopy with an attachment of integral sphere. Photocatalytic characterization of $La_{0.2}Sr_{0.7}Fe_{12}O_{19}$ powders was performed over degradation of Methylene blue (MB) by visible light. A 10.0 ppm methylene blue solution was prepared by putting 0.05g $La_{0.2}Sr_{0.7}Fe_{12}O_{19}$ powders in 50 mL MB solution. The catalytic suspension was firstly stirred in dark for 1 hour, then irradiated under AULIGHT from a 300 W Xenon lamp with a cut filter employed to ensure visible light source (420-700 nm). Photo-degradation of Methylene Blue was monitored in an interval of 30 mins each by measuring the optical absorbance change of the degraded liquids using a Shimazu UV 2100 optical spectrometer.

## 3. Results and Discussion

### 3.1 Structure of $La_{0.2}Sr_{0.7}Fe_{12}O_{19}$

The structure of the as-prepared $La_{0.2}Sr_{0.7}Fe_{12}O_{19}$ powders was detected by an X-ray diffractometer. Figure 1 shows the X-ray diffraction (XRD) pattern of as-prepared $La_{0.2}Sr_{0.7}Fe_{12}O_{19}$ powders; the underneath pink lines (a) are corresponding to the standard diffraction spectrum of pure $SrFe_{12}O_{19}$ (PDF#33-1340).

All the diffraction peaks of the $La_{0.2}Sr_{0.7}Fe_{12}O_{19}$ pattern match well with the standard spectrum of pure $SrFe_{12}O_{19}$, suggesting the formation of pure $La_{0.2}Sr_{0.7}Fe_{12}O_{19}$ compound. There is no second ferrite phase or any other oxide impurities. This diffraction pattern is truly in accordance with the typical formation of diffraction peaks for M-type hexaferrite. The charge has already been balanced by substituting 0.3 $Sr^{2+}$ ions with 0.2 $La^{3+}$ ions. The lattice parameters of $La_{0.2}Sr_{0.7}Fe_{12}O_{19}$ are calculated to be a=5.8724 Å and c=23.014 Å in the hexagonal structure with space group *P63/mmc*. The unit cell volume of $La_{0.2}Sr_{0.7}Fe_{12}O_{19}$, however, has been contracted by 8.76 Å$^3$ (0.59%) in comparison with that of $SrFe_{12}O_{19}$. The overlap of the two diffraction patterns suggests that $La_{0.2}Sr_{0.7}Fe_{12}O_{19}$ shares the same structure with $SrFe_{12}O_{19}$ and 0.2 La atoms have successfully substituted 0.3 Sr atomic sites of the $SrFe_{12}O_{19}$ lattice cells, leaving 10% sites vacant.

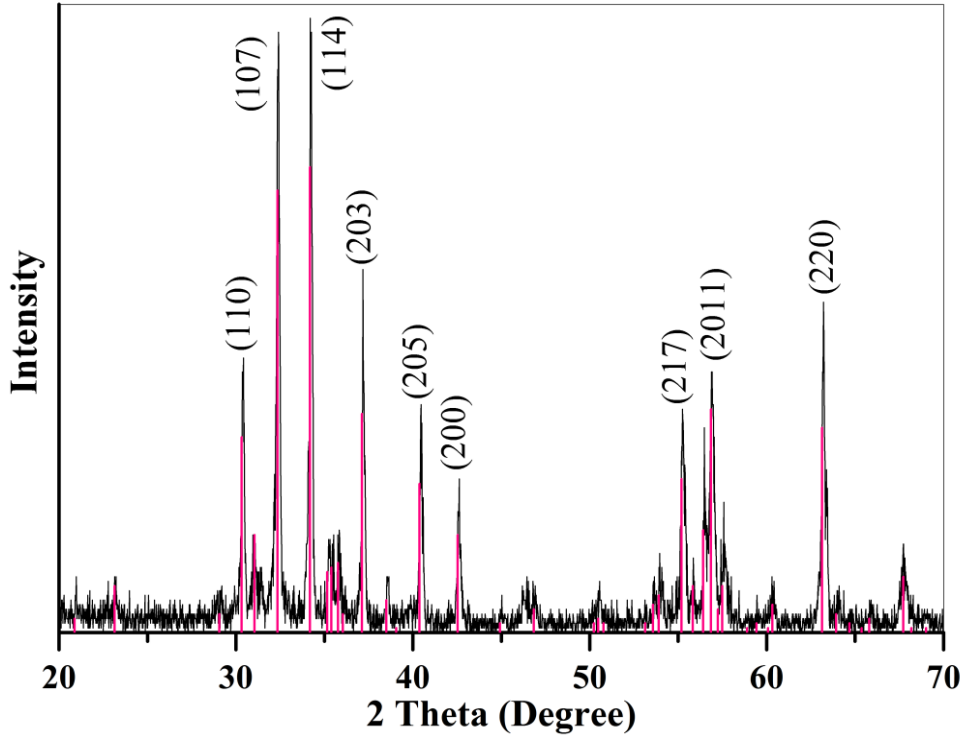

Figure 1: XRD pattern for as-prepared $La_{0.2}Sr_{0.7}Fe_{12}O_{19}$ powders, the underneath pink spectrum represents standard diffraction lines from JCPD #33-1340 card.

## 3.2 Semiconducting Characterization of Magnetic $La_{0.2}Sr_{0.7}Fe_{12}O_{19}$

The as-sintered $La_{0.2}Sr_{0.7}Fe_{12}O_{19}$ powders were pressed into a $\phi 40 \times 3mm$ concave hole on a steel holder, which was attached on an integrated sphere inside a Lambda 750 UV-vis-near infrared spectrometer for optical measurement. Figure 2 shows the UV-VIS-NIR absorption spectrum of as-prepared $La_{0.2}Sr_{0.7}Fe_{12}O_{19}$ powders. The powders were sintered 1100°C for 1 h and subsequently annealed at 800°C in pure $O_2$ atmosphere for three times. The absorption spectrum changes with the change in the difference between photon energy and band gap. The spectrum is composed of 4 segment lines, one within the wavelength region of 250 nm~590 nm, one steeply dropping line within 590 nm~830 nm, one gently sloping line within 830nm~1450nm and the last flat line within 1450nm~2500nm above. Two linear increasing slopes reflect typical semiconductor characterization with indirect band gap. The absorbance is very small when the photon energy is smaller than the band gap energy (λ>1450 nm), since these photons do not have enough energy to excite the valence electrons to make a transition across the band gap into the conduction band. However, when the photon energy is comparable to the band gap energy, the absorbance demonstrates a sudden increase. The majority of the photons within this energy range (590nm<λ<1450nm) were absorbed by the electrons near the top of valence band, those electrons gain the energy from the photons and are then excited to conduction band across the gap, resulting in a sudden rise of the absorbance of the photons. This suggests the absorbance happened in the visible region. When the photon energy is larger than the band gap, the absorption approaches to saturation, therefore the absorbance no longer increases. The peak locating at 590 nm could be assigned to the bounded exiton absorption of $La_{0.2}Sr_{0.7}Fe_{12}O_{19}$.

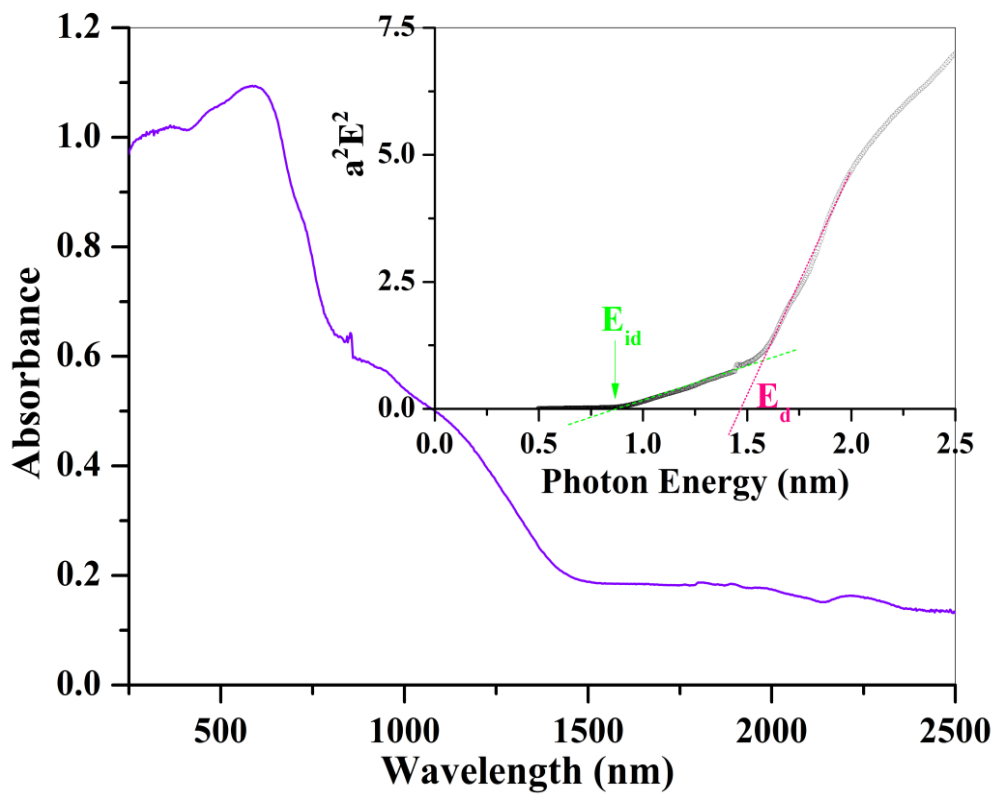

Figure 2: UV-vis-near infrared optical spectrum of $La_{0.2}Sr_{0.7}Fe_{12}O_{19}$ powders, which were calcined at 1100°C for 1 h and subsequent $O_2$ heat treated at 800°C for 3 times.

The absorption data was then plotted as a $\alpha^2 E^2$ versus E, in order to derivate the Tauc plot for determining the band gap energies of semiconducting $La_{0.2}Sr_{0.7}Fe_{12}O_{19}$ compound:

$$(\alpha E)^2 = A(E - E_g) \quad (1)$$

Thus the fundamental process for a Tauc plot is to collect the absorbance data of the samples spanning from below the band gap transition to above it. The inset of Figure 2 displays the Tauc plot of $La_{0.2}Sr_{0.7}Fe_{12}O_{19}$, where the $(\alpha E)^2$ is plotted versus the photon energy E. The occurrence of concave turning point between two sloped segment lines for the second power Tauc plot of $(\alpha E)^2$ versus E confirms the indirect allowed transition, indicating the feature of indirect band gap characterization of $La_{0.2}Sr_{0.7}Fe_{12}O_{19}$. Linearly extrapolating the $a^2E^2$ edges with different slope to zero absorption extracts two lines. The intersection points of the two extrapolating lines with the X-axis are determined to be direct and indirect band gap energies, which are thus estimated to be $E_d=1.47$ eV and $E_{id}=0.88$ eV respectively, as being shown in the inset of Figure 4. Both band gap energies are comparable to that of $PbFe_{12}O_{19}$ [21].

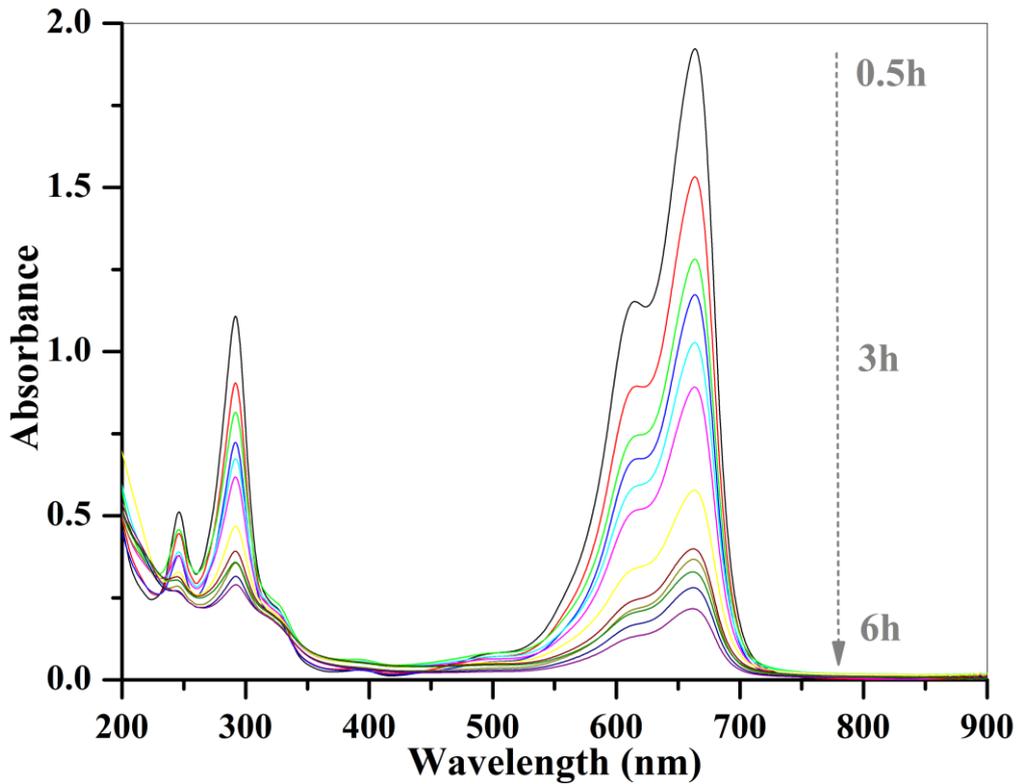

Figure 3: Optical absorbance degradation of aqueous MB by photocatalytic process of $La_{0.2}Sr_{0.7}Fe_{12}O_{19}$ suspension under visible light illumination, the spectrum starts from the illumination duration of 30 minutes and then all the way down to 6 hours with an equal interval of 30 minutes.

Another evidence for semiconducting characteristic of $La_{0.2}Sr_{0.7}Fe_{12}O_{19}$ is the photocatalytic degradation process of organic compounds, which capture photon excited electrons from conduction band of $La_{0.2}Sr_{0.7}Fe_{12}O_{19}$ and are degraded to smaller molecules. Methylene blue (MB) is a representative of a class of dyestuffs resistant to biodegradation.

Here, we choose Methylene blue (MB) as the representative organic compound to check out the photo-degradation performance of $La_{0.2}Sr_{0.7}Fe_{12}O_{19}$, so as to further confirm its semiconducting feature. The degradation was carried out in the aqueous suspension of MB with small amount of $SrFe_{12}O_{19}$ powders under visible illumination.

Figure 3 displays the absorption coefficient decreasing with light illumination time, which reflects photo-degradation process of MB in the presence of La modified M-type strontium hexaferrite ($La_{0.2}Sr_{0.7}Fe_{12}O_{19}$) powders. Under our experimental conditions, there are five maximal absorption peaks of the original MB at 664, 613, 334, 292 and 247 nm. Temporal changes in the concentration of MB were monitored by examining the variations in maximal absorption in UV–visible spectra at 664nm, which is one employed in most literatures [27-29]. Significant temporal concentration degradation of MB was witnessed by the quick descent absorption intensity with the duration of photocatalytic reaction. Figure 4 shows such quick concentration degradation of MB as a function of reaction time upon catalysis of $La_{0.2}Sr_{0.7}Fe_{12}O_{19}$ powders. The concentration of MB degraded exponentially with photocatalytic reaction time. The optical intensity of the main absorption band at 664 nm attenuates down to 0.18 upon 6h photocatalytic reaction with $La_{0.2}Sr_{0.7}Fe_{12}O_{19}$, indicating that degradation of MB has almost completed.

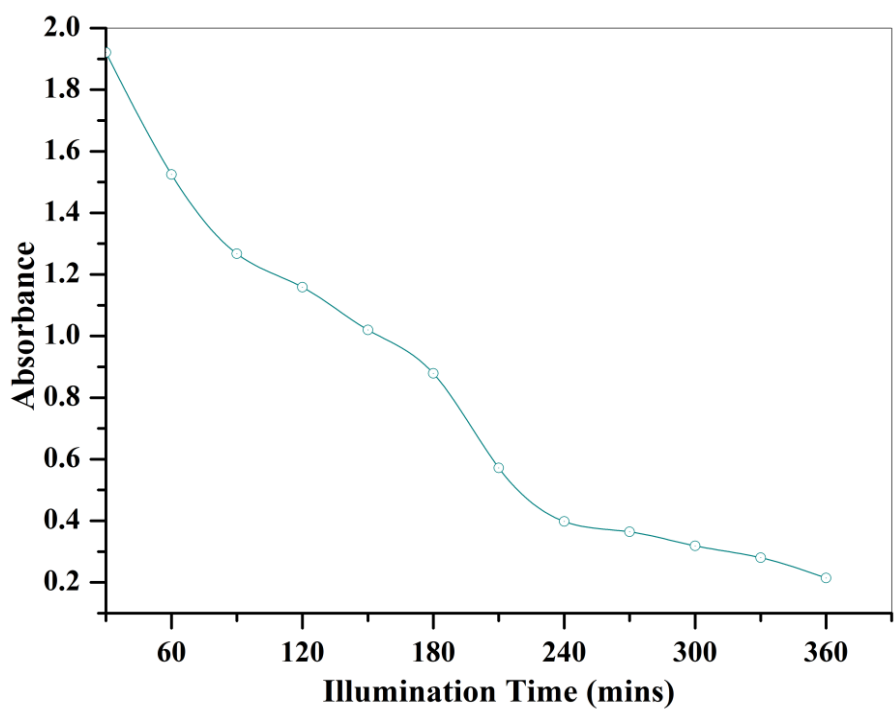

Figure 4：Time-dependent photocatalytic degradation process of MB by $La_{0.2}Sr_{0.7}Fe_{12}O_{19.}$

However, it is pretty interesting to note that the blue shifts of the five spectral bands of MB blue, which appeared during the course of photo-degradation process of MB in the presence of $TiO_2$ and other catalysts[30,31], did not appear in the present optical spectra for photo-degradation process of MB by $La_{0.2}Sr_{0.7}Fe_{12}O_{19}$. It was reported that the spectral band at 664 nm blue-shifts by as much as 54 nm from 664 to 610 nm during the course of the MB photo-degradation by $TiO_2$[32]. These kinds of blue shifts were attributed to different steps of *N*-demethylation of MB during its photocatalytic degradation process[32]. In other words, MB was degraded into the final product in a multiple-steps manner, including *N*-demethylation, deamination and oxidative degradation in case of photocatalytic process of $TiO_2$. The *N*-demethylated intermediates differ their maximum absorption peaks from MB and thus caused stepwise blue shifts[32]. When $La_{0.2}Sr_{0.7}Fe_{12}O_{19}$ was employed as photocatalyst for degradation of MB, the five absorption bands did not show any blue shifts with irradiation time, but the intensity decreases rapidly, as being shown in Figure 3. The decrease in peak intensity with increase in time duration is attributed towards the conversion of pure methylene blue to a photo-catalytically degraded inorganic compound of the same molecule. Non-blue shifts of the absorption bands indicate that no intermediates were formed during the course of photocatalytic degradation process of MB by $La_{0.2}Sr_{0.7}Fe_{12}O_{19}$, the degradation process could complete in just one stepwise.

### 3.3 Magnetic Properties of La substituted $SrFe_{12}O_{19}$ Compound

The magnetic properties of $La_{0.2}Sr_{0.7}Fe_{12}O_{19}$ powders were measured by a Physical Property Measurement System (PPMS) at room temperature. Figure 5 displays the ferromagnetic hysteresis M-H loop of $O_2$ annealed $La_{0.2}Sr_{0.7}Fe_{12}O_{19}$. The magnetic properties of $La_{0.2}Sr_{0.7}Fe_{12}O_{19}$ have been greatly improved in comparison with that of pure $SrFe_{12}O_{19}$.[33] The remnant magnetic moment (M) and coercive field ($H_C$) of $La_{0.2}Sr_{0.7}Fe_{12}O_{19}$ compound are determined to be 52 emu/g and 5876 Oe, respectively. The coercive field ($H_C$) is almost the same as pure $SrFe_{12}O_{19}$, but the remnant magnetic moment ($M_r$) of $La_{0.2}Sr_{0.7}Fe_{12}O_{19}$ has been enhanced by around 69.4% in comparison with that of pure $SrFe_{12}O_{19}$, since the substituting $La^{3+}$ provides one more unpaired electron spin. Similar improvement of the magnetic polarization was also observed in La doped $PbFe_{12}O_{19}$.[34]

Sintering the specimen inside an oxygen deficient sealed air furnace would generate many $Fe^{2+}$ ions and thus could deteriorate the magnetic polarisation ability of $La_{0.2}Sr_{0.7}Fe_{12}O_{19}$. Therefore, converting $Fe^{2+}$ to $Fe^{3+}$ during annealing in oxygen environment plays an important role on improving the magnetic performance of $La_{0.2}Sr_{0.7}Fe_{12}O_{19}$, since $Fe^{3+}$ could provide one more unpaired electron spin, which attributes toward larger magnetic polarisation in hexaferrite. With both better magnetic properties and excellent semiconducting feature, $La_{0.2}Sr_{0.7}Fe_{12}O_{19}$ is a kind of typical multifunctional material, which simultaneously integrates several functions of magnetics and optics into one structure. The combination of multiple functions in one phase could be able to trig generation of new photocatalytic applications in environmental field in virtue of its magnetic semiconducting feature.

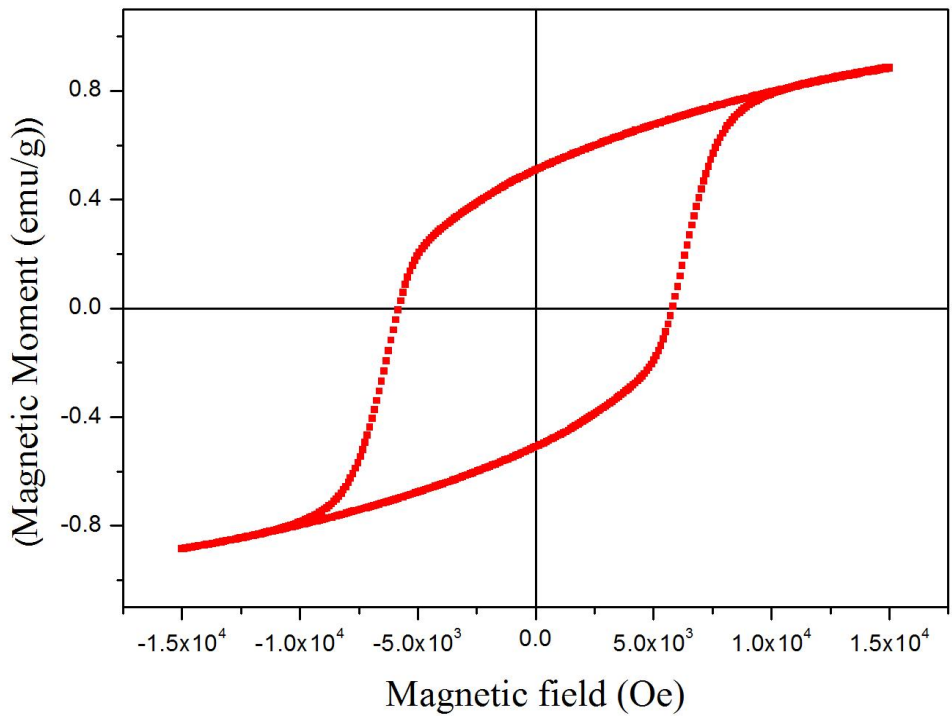

Figure 5: Magnetic hysteresis loop of $La_{0.2}Sr_{0.7}Fe_{12}O_{19}$ powders being sintered at 1100 °C for 1 h and subsequently annealed in pure $O_2$ for 9 hours.

Magnetic semiconductor powders have great advantage over the conventional catalytic ones, since they could be made into magnetic fluid together with the organic dye molecules, which could be easily brought into magnetic suspension in water and greatly enhance the area of the catalysts exposing to the sunlight, as such the photocatalytic reaction efficient could be

## 4. Conclusion

La modified strontium M-type hexaferrite (La$_{0.2}$Sr$_{0.7}$Fe$_{12}$O$_{19}$), an M-type hexaferrite compound, was prepared by a polymer precursor method. The compound shows stronger magnetic performance than pure SrFe$_{12}$O$_{19}$, the remnant magnetic moment (M$_r$) and coercive field (H$_c$) reaches as high as 52 emu/g and 5876 Oe, respectively. In addition, it also demonstrates semiconducting feature, which was revealed by UV-Vis optical spectroscopy. The direct and indirect band gap energies were determined to be 1.47 eV and 0.88 eV, respectively. The photocatalytic degradation of MB has been tested on the aqueous dispersions containing La$_{0.2}$Sr$_{0.7}$Fe$_{12}$O$_{19}$ powders under visible light illumination. Gradually with the increase in the duration of photodegradation reaction, the colour of the MB dispersion solution changes from deep blue to pale white, while the absorption intensity reduces rapidly from around 1.80 down to 0.18. Hypsochromic effects resulting from *N*-demethylation of MB, which presented in case of photo-degradation of MB on TiO$_2$ and other conventional catalysts, did not appear in this case. But oxidative degradation of MB occurred during irradiation, suggesting that the photo-degradation of MB into final product completes in just one stepwise.

**Acknowledgement:**

The authors appreciate Dr Xiaoguang Huang for the support of optical spectrum and photocatalytic measurements. The authors acknowledges the financial support from the National Natural Science Foundation of China under grant No. 11774276.